\documentclass[twocolumn,showpacs,preprintnumbers,amsmath,amssymb]{revtex4}
\usepackage{graphicx}
\usepackage{bm}
\usepackage{dcolumn}

\def\prl#1#2#3{{ Phys.   Rev.   Lett.  } {\bf #1}, #2 (#3)}

\def\pre#1#2#3{Phys.   Rev.   E {\bf #1}, #2 (#3)}

\def\pra#1#2#3{Phys.   Rev.   A {\bf #1}, #2 (#3)}

\def\epl#1#2#3{Europhys. Lett. {\bf #1}, #2 (#3)}
\def\jfm#1#2#3{J. Fluid Mech. {\bf #1}, #2 (#3)}
\def\jscc#1#2#3{J. Sci. Comp. {\bf#1}, #2 (#3)}
\def\pfa#1#2#3{Phys. Fluids A {\bf#1}, #2 (#3)}
\def\jap#1#2#3{J. Appl. Phys. {\bf#1}. #2 (#3)}
\def\arfm#1#2#3{Ann. Rev. Fluid Mech. {\bf#1}, #2 (#3)}

\def\noi{\noindent}
\def\bc{\begin{center}}
\def\ec{\end{center}}
 \newcommand{\bea}{\begin{equation}}
 \newcommand{\eea}{\end{equation}\noi}
 \newcommand{\ber}{\begin{eqnarray}}
 \newcommand{\eer}{\end{eqnarray}\noi}
\begin{document}
\title{Growth Models and Models of Turbulence : A Stochastic Quantization 
Perspective}
\author{Himadri   S.   Samanta}\email{tphss@mahendra.iacs.res.in}
\author{J.   K. Bhattacharjee}\email{tpjkb@mahendra.iacs.res.in}
\affiliation{Department of Theoretical Physics,
Indian Association for the Cultivation of Science \\
Jadavpur, Calcutta 700 032, India}
\author{D. Gangopadhyay}\email{debashis@boson.bose.res.in} 
\affiliation{S.N.Bose National Centre For Basic Sciences, JD-Block,
Sector-III, Salt Lake City, Kolkata-700098, India }
\date{\today}
\begin{abstract}
We consider a class of growth models and models of turbulence
based on the randomly stirred fluid. The similarity between the predictions
of these models, noted a decade earlier, is understood on the basis of a 
stochastic quantization scheme.

\end{abstract}

\maketitle

Over the last decade, it has been noticed that two different problems
in non equilibrium statistical mechanics seem to have the same 
characteristics. One is the growth of interfaces under random deposition -
the sort of growth that characterises molecular beam epitaxy 
and the other is the behaviour of velocity in a randomly stirred fluid, 
which arguably is a decent model of turbulence. The surface deposition 
model has its origin in the works of Wolf and Villain\cite{1} and Das Sarma and
Tamborenea\cite{2}. These were discrete models where particles were dropped at 
random on an underlying lattice. Having reached the surface, the 
particle diffuses along the surface until it finds an adequate number
of nearest neighbours and then settles down. The surface is characterised by
the correlation of height fluctuations. The equal time correlation 
of the height variable $h(\vec{r},t)$ is expressed as

\bea\label{eqn1}
<[\Delta h(\vec{r},t)]^{2}>=<[h(\vec{r}+\vec{r_{0}},t)-h(\vec{r}_{0},t)]^{2}>
\propto  r^{2\alpha}
\eea

as $t$ becomes longer than $L^{z}$, where $L$ is the system size.
In the above, $\alpha $ is called the roughness exponent and $z$
the dynamic exponent. The surface is rough if $\alpha $ is positive.
The continuum version of this model was introduced by Lai and Das Sarma\cite{3}.
The similarity of this growth model with models of fluid turbulence
was first noticed by Krug\cite{4} who studied the higher order correlation 
functions and found 
\bea\label{eqn2}
<[\Delta h(\vec{r})]^{2p}> \propto r^{2\alpha_{p}}
\eea

where $\alpha_{p}$ is not a linear function of $p$. This is a 
manifestation of intermittency - the single most important feature
of fluid turbulence\cite{5,6,7,8}. It says that the distribution function for 
$\mid \Delta h(\vec{r},t)\mid$ has a tail which is more long lived 
than that of a Gaussian. The relevant quantity in fluid turbulence is 
the velocity increment $\Delta v_{\alpha}=v_{\alpha}(\vec{r}+\vec{r}_{0},t)
-v_{\alpha }(\vec{r}_{0},t)$, where we are focussing on the 
$\alpha ^{th}$ component. It is well known that $\Delta v_{\alpha}$ 
exhibits multifractal characteristics or intermittency and the multifractal 
exponents have been measured quite carefully\cite{9}. The multifractal nature
of the velocity field is expressed by an equation very similar to 
Eq.(\ref{eqn2}),
\bea\label{eqn3}
<\mid \Delta v_{\alpha}(\vec{r},t)\mid ^{2p}> \propto r^{\zeta_{p}}
\eea 
 
where $\zeta _{p}$ is not a linear function of $p$. From some considerations
of singular structures, She and Leveque\cite{10} arrived at the formula
$\zeta_{p}=\frac{2p}{9}+2[1-(\frac{2}{3})^{2p/3}]$, 
which gives a very reasonable account of the experimentally
determined multifractal indices. It is the similarity between the growth model
and the turbulence characteristics that is interesting.

The randomly stirred Navier Stokes equation\cite{11,12,13,14,15} 
in momentum space reads

\bea\label{eqn4}
\dot{v}_{\alpha}(k)+i k_{\beta}\sum_{\vec{p}}v_{\beta}(\vec{p})
v_{\alpha}(\vec{k}-\vec{p})= i k_{\alpha}P - \nu k^{2}v_{\alpha}+f_{\alpha}
\eea

with the incompressibility condition appearing as 
$k_{\alpha}v_{\alpha}(k)=0$. The random force $f_{\alpha}$ has the equal 
time correlation 
\bea\label{eqn5}
<f_{\alpha}(\vec{k}_{1})f_{\beta}(\vec{k}_{2})>=
2\frac{D_{0}}{k_{1}^{D-4+y}}\delta_{\alpha \beta} \delta (\vec{k}_{1}+
\vec{k}_{2})
\eea

where $D$ is the dimensionality of space and $y$ is a parameter 
supposed to give the fully developed turbulence for $y=4$. In this 
model, the pressure term does not qualitatively alter the physics
of turbulence and hence it is often useful to study the pressure free
model. This was extensively done by Checklov and Yakhot\cite{16} and Hayot and 
Jayaprakash\cite{17} in $D=1$. If we write $V_{\alpha}=\partial _{\alpha}h$,
then
\bea\label{eqn6}
\dot{h}(k)=-\nu k^{2}h(k)-1/2 \sum \vec{p}\cdot (\vec{k}-\vec{p})
h(\vec{p})h(\vec{k}-\vec{p})+g(k)
\eea

where

\bea\label{eqn7}
<g_{\alpha}(\vec{k})g_{\beta}(\vec{k}^{\prime})>
= \frac{2D_{0}}{k^{D-2+y}}\delta(\vec{k}+\vec{k}^{\prime})
=\frac{2D_{0}}{k^{2\rho}}\delta(\vec{k}+\vec{k}^{\prime})
\eea

This is the Medina, Hwa and Kardar model\cite{18} in $D$-dimensions.
Turning to the growth model, the linear equation for surface 
diffusion limited growth is the Mullins - Sereska model\cite{19} given by
\bea\label{eqn8}
\frac{\partial h(k)}{\partial t} = -\nu k^{4}h(k) + \eta (k)
\eea

where, $<\eta(\vec{k})\eta(\vec{k}^{\prime})>=2D_{0}\delta (\vec{k}
+\vec{k}^{\prime})$ and its generalisation to include nonlinear 
effects is the Lai-Das Sarma model\cite{3} defined as

\bea\label{eqn9}
\frac{\partial h(\vec{k})}{\partial t} = -\nu k^{4}h(\vec{k})-
\frac{\lambda}{2}k^{2} \sum \vec{p}\cdot(\vec{k}-\vec{p})
h(\vec{p})h(\vec{k}-\vec{p})+\eta(\vec{k})
\eea

The various properties of the model have been very well
studied\cite{20,21,22,23,24}.

Our focus is on the similarity between the growth model and the 
model of turbulence. In some sence, these two widely different models
(one with coloured noise, the other with white noise ) have to be related.
We introduce a technique of handling non equilibrium problems that
is based on stochastic quantization\cite{25} and show that the two models can be 
made to look quite similar. We consider a scalar field $\phi(\vec{k})$
in momentum space satisfying the equation of motion
\bea\label{eqn10}
\dot{\phi}(\vec{k}) = -L(\vec{k})\phi (\vec{k})-M(\phi) + g(\vec{k})
\eea

$M(\phi)$ is a non linear term in $\phi$ and $g(k) $ is a noise
which can be coloured in general and we will take it to be of the 
form of Eq.(\ref{eqn7}). The probability distribution corresponding 
to the noise is given by

\bea\label{eq11}
P(g) \propto exp - \int \frac{d^{D}k}{(2\pi)^{D}}
\frac{dw}{2\pi}\frac{k^{2\rho}}{4D_{0}}
g(k,w) g(-k,-w)
\eea

In momentum-frequency space, Eq.(\ref{eqn10}) reads

\bea\label{eq12}
[-iw + L(k)]\phi(k,w)+M_{k,w}(\phi) = g(k,w)
\eea
 
The probability distribution written in terms of $\phi(k,w)$
instead of $g(k,w)$ is
\ber\label{eq13}
 P &\propto & exp\{-\frac{1}{4D_{0}}\int \frac{d^{D}k}{(2\pi)^{D}}
\frac{dw}{2\pi}k^{2\rho}\nonumber \\ & &
\{[-iw+L(k)]\phi(k,w)+M_{k,w}(\phi)\}\nonumber \\ & &
\{[iw+L(-k)]\phi(-k,-w)+M_{-k,-w}(\phi)\}\}\nonumber\\ 
&=&\int \mathcal{D}[\phi]e^{-\frac{1}{4D_{0}}\int 
\frac{d^{D}k}{(2\pi)^{D}}\frac{dw}{2\pi}S(k,w)}\nonumber\\& &
\eer

At this point of development, the usual practice is to 
introduce a response field $\tilde{\phi}$, work out the 
response function as $<\phi \tilde {\phi}>$ and the correlation
function as $<\phi \phi>$. There is no fluctuation dissipation theorem
to relate the two and hence two independent considerations 
are necessary. We now exploit the stochastic quantization scheme 
of Parisi and Wu to introduce a fictitious time $'\tau'$ and consider 
all variables to be functions of $\tau$ in addition to $\vec{k}$
and $w$. A Langevin equation in $'\tau'$ space as
\bea\label{eq14}
\frac{\partial \phi(k,w,\tau)}{\partial \tau}=
- \frac{\delta S}{\delta \phi(-k,-w,\tau)}
+n(k,w,\tau)
\eea
with $<nn>=2\delta(\vec{k}+\vec{k}^{\prime})\delta(w+w^{\prime})
\delta(\tau -\tau^{\prime})$.

This ensures that as $\tau \rightarrow \infty$, the distribution 
function will be given by $S(k,w)$ of Eq.(\ref{eq13}) and in the  
$\tau$-space ensures a fluctuation dissipation theorem. From Eq.(\ref{eq13}),
we find the form of Langevin equation to be 
\ber\label{eq15}
& & \frac{\partial \phi(k,w,\tau)}{\partial \tau}=
k^{2\rho}(\frac{w^{2}+L^{2}}{2D_{0}})\phi(k,w,\tau)
-\frac{\delta}{\delta \phi}[\int \frac{d^{D}p}{(2\pi)^{D}}
\frac{dw^{\prime}}{2\pi}\nonumber \\ & &
p^{2\rho} \{(-iw^{\prime}+L(p))\phi(\vec{p},w^{\prime})
M_{-\vec{p},-w^{\prime}}(\phi)\nonumber \\& &+
(iw^{\prime}+L(-p))
\phi(-\vec{p},w^{\prime})M_{\vec{p},w^{\prime}}\}]\nonumber \\ & &
-\frac{\delta}{\delta \phi}[\int \frac{d^{D}p}{(2\pi)^{D}}
\frac{dw^{\prime}}{2\pi} p^{2\rho}M_{\vec{p},w^{\prime}}(\phi)
M_{-\vec{p},-w^{\prime}}(\phi)] +n(\vec{k},w,\tau)\nonumber \\ & &
\eer

The correlation functions calculated from the above Langevin equation
lead to the correlation functions of the original model as 
$\tau \rightarrow \infty$. For proving scaling laws and noting equivalences,
it suffices to work at arbitrary $\tau$. It is obvious from Eq.(\ref{eq15})
that in the absence of the nonlinear terms (the terms involving $M(\phi)$), 
the Greens function $G^{(0)}$ is given by 
\bea\label{eq16}
[G^{(0)}]^{-1}= -i\Omega_{\tau} +k^{2\rho}\frac{w^{2}+L^{2}}{2D_{0}}
\eea
where $\Omega_{\tau}$ is the frequency corresponding to the fictitious
time $\tau$. As is usual, the effect of the nonlinear terms, leads to the
Dysons' equation 
\bea\label{eq17}
G^{-1}=[G^{(0)}]^{-1} +\sum (k,w,\Omega_{\tau})
\eea

The correlation function is given by the fluctuation dissipation 
theorem as $C=\frac{1}{\Omega_{\tau}}Im G$.
Let us start with Eq.(\ref{eqn6}) which relates to fluid turbulence.
The linear part of the corresponding Eq.(\ref{eq16}) gives
\bea\label{eq18}
[G^{(0)}]^{-1} = -i\Omega_{\tau} + \frac{k^{2\rho}}{2D_{0}}
(w^{2}+\nu^{2}k^{4})
\eea

The $\tau \rightarrow \infty$ limit of the equal time correlation function
is $2D_{0}/k^{2\rho}(w^{2}+\nu^{2}k^{4})$, which leads, in $D=1$, to
$\alpha =(1+2\rho)/2$. The dynamic exponent is clearly $z=2$. If we now turn
to the growth model of Eq.(\ref{eqn9}) and consider the linear part of the 
relevant form of Eq.(\ref{eq15}), then $[G^{(0)}]^{-1}=-i\Omega_{\tau}
+(w^{2}+\nu^{2}k^{8})/2D_{0}$ and the corresponding $\alpha = 3/2$ in
$D=1$. The dynamic exponent $z$ is $4$. We note that although the dynamic 
exponents never match, the two roughness exponents are equal for $\rho =1$.
This is what is significant. 

We now turn to the nonlinear terms and treat them to one loop order. 
For Eq.(\ref{eqn6}), $M_{k,w}(\phi)=(1/2)\sum_{\vec{p}}\vec{p}\cdot 
(\vec{k}-\vec{p})\phi(p)\phi(\vec{k}-\vec{p}) $, while for Eq.(\ref{eqn9})
$M_{k,w}(\phi)=(1/2) k^{2}\sum_{\vec{p}}\vec{p}\cdot( \vec{k}-\vec{p})
\phi(\vec{p})\phi(\vec{k}-\vec{p})$. 
The nonlinear term which involves two $M$'s in Eq.(\ref{eq15}) gives 
a one loop correction which is independent of external momenta and 
frequency and hence is not relevent at this order. It is the term involving one
$M$ which is important and for Eq.(\ref{eqn6}), this has the structure 
$k^{2\rho}(-iw+\nu k^{2})(\vec{p}\cdot(\vec{k}-\vec{p}))\phi(\vec{p})
\phi(\vec{k}-\vec{p})$. For Eq.(\ref{eqn10}), the corresponding structure is
$(-iw+\nu k^{2})[k^{2}\vec{p}\cdot(\vec{k}-\vec{p})]\phi(\vec{p})
\phi(\vec{k}-\vec{p})$. For $\rho=1$, the two nonlinear terms have very similar
structure! The scaling of the correlation function determines the 
roughness exponent. Now the one loop graph in both cases are composed of two 
vertices, one response function and one correlation function. 
While the dynamic exponent $z$ will differ the momentum count of the one loop
graph for the fluid must agree with that for the interface growth since for
$\rho =1$, the vertex factor agree, the correlation functions tally and 
the frequency integrals of $G^{(0)}$ match. 
Thus at $\rho=1$, the perturbation theoretic evaluation of $\alpha$
for for the two models will be equal.

How big is $\alpha$ in the growth model? A one loop self consistent calculation yields the answer in a trivial fashion. The structure of $\sum$ is 
$\int d^{D}p dw d\Omega_{\tau}V.V.GC$. We recall that $C$ is $1/\Omega_{\tau}
ImG$ and hence dimensionally this is $\int d^{D}pdw VVGG$. If the 
frequency scale is to be modified to $k^{z}$ from $k^{4}$ with $z<4$, 
then $G$ scales as $k^{-2z}$ and $V~k^{4+z}$ and hence $\sum ~ k^{D+8-z}$
which has to match $k^{2z}$. This yields $z=(D+8)/3$.

A ward identity shows $\alpha+z=4$ and thus $\alpha=(4-D)/3$.
At $D=1$, $\alpha=1$ and matches the $\rho=1$ results of $\alpha=1$
for the fluid model. But as is apparent from the work of referrence(17)
, this is where the multifractal nature sets in for the fluid because of the 
nonlinear term. The identical structure of the growth model nonlinearity
tells us that in $D=1$, it too will have multifractal behaviour.
Thus, we see that the growth model and the turbulence model are not 
in the same universality class since the dynamic exponents are different but
the structure of the Langevin equation in the fictitious time makes 
it clear that they will have the same roughness behaviour.

\end{document}